\documentclass[doublespacing]{elsart}

\usepackage{graphicx}
\usepackage{amssymb}
\def\tele{Sr$_{14}$Cu$_{24}$O$_{41}$}
\def\cgo{CuGeO$_{3}$}
\def\scbo{SrCu$_2$(BO$_3$)$_2$}

\begin{document}
\begin{frontmatter}

\journal{SCES'2001: Version 2}

\title{Heat transport in SrCu$_2$(BO$_3$)$_2$ and CuGeO$_3$}

\author[col]{M. Hofmann}
\author[col]{T. Lorenz}
\author[col]{A. Freimuth}
\author[theo]{G.S. Uhrig}
\author[tok]{H. Kageyama}
\author[tok]{Y. Ueda}
\author[par]{G. Dhalenne}
\author[par]{A. Revcolevschi}

\address[col]{II. Phys. Institut, Universit\"{a}t zu K\"{o}ln,
 Z\"{u}lpicher Str. 77, 50937 K\"{o}ln, Germany}
\address[theo]{Inst. f\"{u}r Theoretische Physik, Univ. zu K\"{o}ln,
 Z\"{u}lpicher Str. 77, 50937 K\"{o}ln}
 \address[tok]{Inst. for Sol. State Phys., Univ. of Tokyo,
  Kashiwanoha 5-1-5, Kashiwa,  Japan}
\address[par]{Lab. de Physico-Chimie des Solides, Univ.
 Paris-Sud, 91405 Orsay, France}


\begin{abstract}

In the low dimensional spin systems SrCu$_2$(BO$_3$)$_2$ and
CuGeO$_3$ the thermal conductivities along different crystal directions
show pronounced double-peak structures and strongly depend on magnetic fields. For SrCu$_2$(BO$_3$)$_2$
the experimental data can be described by a purely phononic
heat current and resonant scattering of phonons by magnetic excitations. A similar effect seems to be important in \cgo , too
but, in addition, a magnetic contribution to the heat transport may be present.

\end{abstract}


\end{frontmatter}

One topic of interest in low dimensional~(D) quantum spin systems is the dynamics of magnetic excitations, which
e.g. are expected to move without dissipation in an S=1/2~Heisenberg chain~\cite{zotos97}.
This may lead to an unusual thermal conductivity in
real, (quasi-)low-D systems. Experimental evidence
for a large magnetic contribution $\kappa^{mag}$ to the heat transport has been found in
\tele\ containing both, spin chains and ladders~\cite{solog00}.
Here, $\kappa$ shows an enormous maximum around 200~K which is present {\em only} for
heat transport parallel to the ladder (and chain) direction.  This has been interpreted as a huge
magnetic contribution to $\kappa$, which adds to a phononic contribution $\kappa^{ph}$ giving rise
to low-T maxima around 20~K for all three crystal directions.
In this paper we present a study of the anisotropic heat transport
in the spin Peierls (SP) system \cgo\ and in the 2D dimer spin system \scbo ~\cite{kagey00a},
whose magnetic planes represent a realization of the Shastry Sutherland model~\cite{shast81b}.

\begin{figure}
     \centering
     \includegraphics[width=10cm,clip]{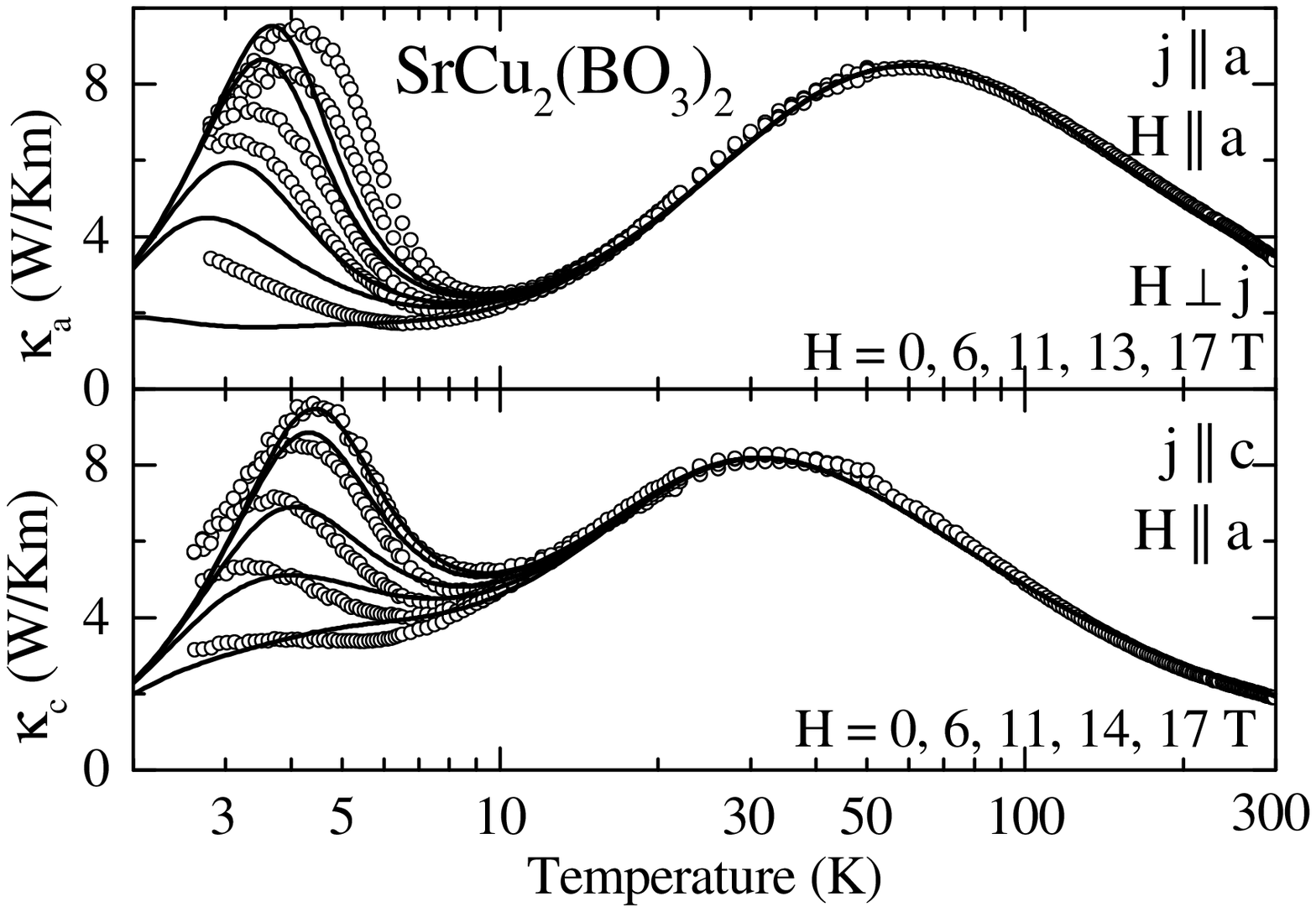}
     \caption{Thermal conductivity $\kappa_a$ (top) and $\kappa_c$
 (bottom) of tetragonal \scbo\ on a logarithmic temperature scale for various fields.  On increasing field the
 low-$T$ maximum is suppressed. Lines are theoretical curves calculated for the same fields.}
     \label{sc}
 \end{figure}

The thermal conductivities of tetragonal \scbo\ parallel ($\kappa_c$) and perpendicular ($\kappa_a$) to the magnetic planes show pronounced
double peaks with a first maximum around 4~K for both directions and a second one around 60~K and 30~K for $\kappa_a$ and
$\kappa_c$, respectively (see Fig.~\ref{sc}). In a magnetic field the maxima at low T are suppressed whereas those at
high T remain unchanged. At first glance, one might suspect that the field-dependent double peaks result from the superposition of a
phononic and a magnetic contribution. However, the similar behavior of $\kappa_a$ and $\kappa_c$ makes any explanation depending on
a $\kappa^{mag}$ very unlikely, since $\kappa^{mag}$ perpendicular to the magnetic planes should be negligible. Moreover, the
triplet excitations of \scbo\ are almost dispersionless along the magnetic planes~\cite{kagey00a}, so that even within the planes
$\kappa^{mag}$ is expected to be very small.
This led us to a model that is based on a purely phononic heat current and explains the field-dependent double peaks
by resonant scattering of phonons by magnetic excitations. It can be visualized as
(i) the absorption of a phonon of energy $\omega$ by exciting a (magnetic) two-level system and (ii)
its subsequent de-excitation by emitting another phonon of same energy but
different momentum (for more details see~\cite{hofmann01a}). Such a scattering is most active in a certain
temperature range depending on the energy splitting of the two-level system.
Spin conservation requires a thermally excited triplet and the phonon excites a second triplet
which combines with the first one to total spin $S_{tot}=1$.
Thus, the scattering rate depends on the thermal population of excited triplets. It also depends on the magnetic field
due to the Zeeman splitting. Within this model
our experimental data for zero field can be fitted very well. Since these fits fix all parameters the data for higher fields
are calculated from the Zeeman splitting without further parameter adjustment.
As shown by the solid lines in Fig.~\ref{sc} our model describes the general features of $\kappa_a$ and $\kappa_c$ very well;
$\kappa_c$ is even quantitatively reproduced up to 17~T, whereas the field dependence of $\kappa_a$ is slightly
overestimated.

\begin{figure}
     \centering
     \includegraphics[width=10cm,clip]{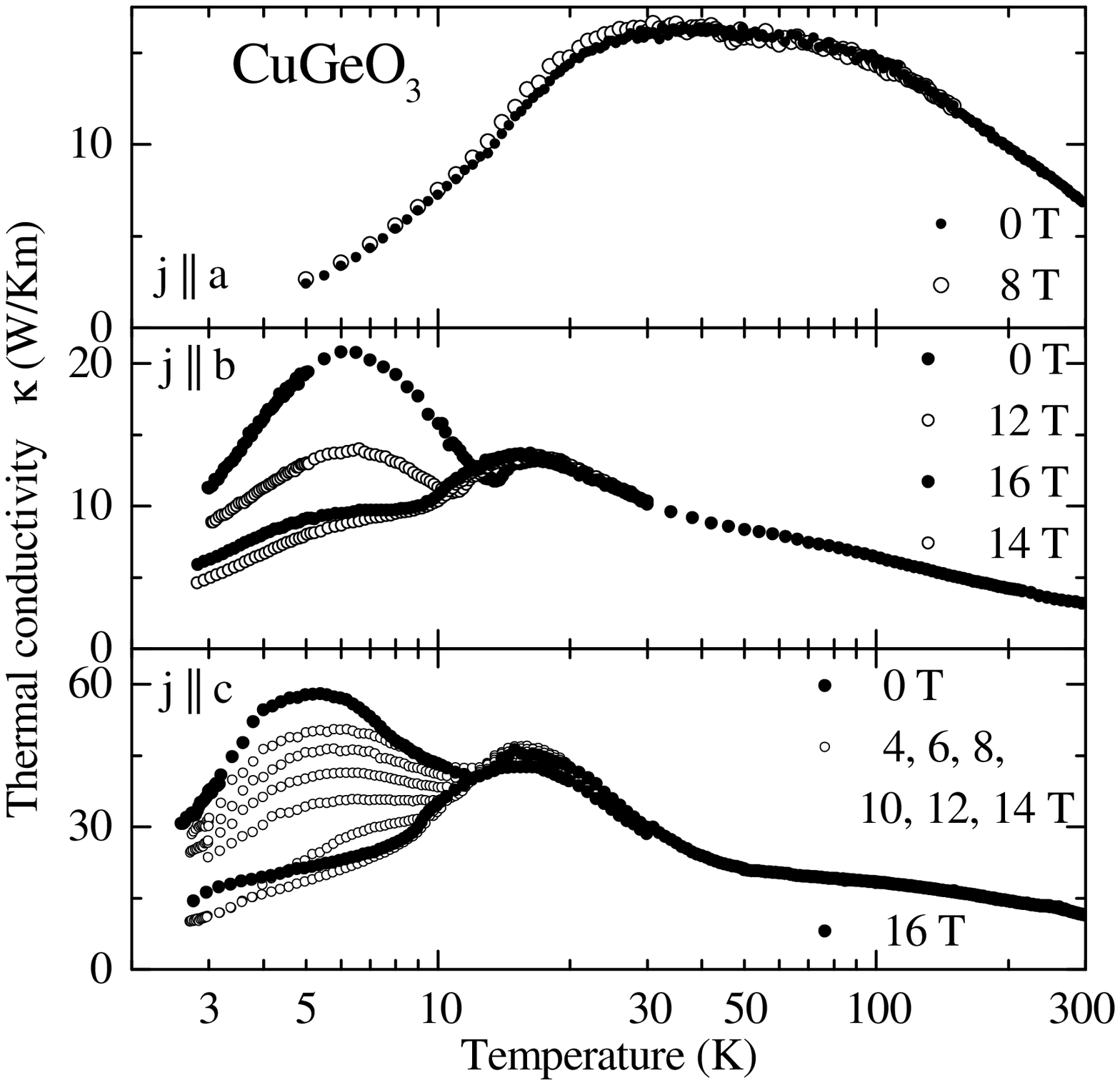}
     \caption{Thermal conductivity $\kappa_a$, $\kappa_b$ and $\kappa_c$ (from top to bottom)
     of orthorhombic \cgo\ on a logarithmic temperature scale for various fields.
     $\kappa_a$ is field-independent whereas $\kappa_b$ and $\kappa_c$ decrease with increasing field up
     to 14~T and then increase again. }
     \label{cg}
 \end{figure}

The thermal conductivities along the $a$, $b$, and $c$ axis of orthorhombic \cgo\ are shown in Fig.~\ref{cg}.
Our results confirm those previously measured for $\kappa_c$ ($||$ to the spin chains)~\cite{ando} showing one maximum
below and a second one above the SP transition at $T_{\rm SP}\sim 14$~K. With increasing field
the low T maximum is continuously suppressed up to 14~T and then slightly increases again. This non-monotonic field dependence
is related to the occurrence of the so called incommensurate phase above $\sim 12.5$~T~\cite{lorenz}.
Although the data of \cgo\ qualitatively resemble those of \scbo\
the situation in the two compounds is rather different.
A sizeable $\kappa^{mag}$ --- though not necessary for an explanation of a field-dependent
double peak --- may be present in \cgo\ since the triplet excitations have considerable dispersion~\cite{INS}.
In addition, there is the SP transition. Above $T_{\rm SP}$ the energy gap
closes and strong structural fluctuations are observed up to $\sim 35$~K~\cite{schoeffel96a}.
All these effects may influence $\kappa^{ph}$ and/or $\kappa^{mag}$ making a reliable
description of $\kappa$ difficult.

In order to clarify whether a magnetic contribution to $\kappa_c$ is present in \cgo\ we have also measured
$\kappa_b$ and $\kappa_a$ ($\perp$ to the spin chains). As shown in Fig.~\ref{cg} $\kappa_b$ behaves very similar to
$\kappa_c$ apart from the absolute value which is a factor of $\sim 3$ smaller for  $\kappa_b$ than for  $\kappa_c$.
An anisotropy ratio of $\sim 3$ is present between the sound velocities along $c$ and $b$
($v_c\sim 7600$~m/s and $v_b\sim 2400$~m/s~\cite{saintpaul95a}), but also for the triplet dispersions
along $c$ and $b$~\cite{INS}. Therefore, one can neither exclude nor confirm the presence of a sizeable
magnetic contribution to the heat current from this anisotropy.
Finally, we consider the $a$ axis. The sound velocity amounts to $v_a\sim 3600$~m/s~\cite{saintpaul95a} whereas
the magnetic dispersion is essentially zero~\cite{INS}, i.e. $\kappa^{mag}_a$ is negligible.
At 300~K the anisotropy between  $\kappa_c$,  $\kappa_b$ and (the purely phononic) $\kappa_a$ reflects that of the
$v_i$ ($i=a,b,c$). Therefore $\kappa^{mag}_c$ and $\kappa^{mag}_b$ can also be neglected at room temperature. This is not
surprising because the magnetic coupling constants in \cgo\ are much smaller; $J_c\sim 120 -180$~K and $J_b\sim 0.1 J_c$ depending
on the model~\cite{INS,fabri}.
Unfortunately, $\kappa_a$ cannot help to separate $\kappa^{ph}$ and $\kappa^{mag}$ along $b$ and $c$ at low T because
the layered structure of \cgo\ along $a$ limits the mean free path by so-called sheet-like faults.
This prevents a strong increase of $\kappa_a$ at low T and already explains the completely different behavior of $\kappa_a$ compared
to $\kappa_b$ and $\kappa_c$, regardless their field dependent double-peaks.

In conclusion, the thermal conductivities of \cgo\ and \scbo\ show pronounced, magnetic-field dependent double-peaks.
For \scbo\ this arises from resonant scattering of a phononic heat current by magnetic excitations.
In \cgo\ a similar effect seems to play a role but further studies are necessary in order to clarify the dynamics
of the magnetic excitations.

\newpage

\end{document}